\pdfoutput=1
\RequirePackage{ifpdf}
\ifpdf 
\documentclass[pdftex]{sigma}
\else
\documentclass{sigma}
\fi

\def\tl{\tilde}

\def\gm{\gamma}
\def\lm{\lambda}
\def\AG{A_{\emph{G}}}
\def\BG{B_{\emph{G}}}
\def\CG{C_{\emph{G}}}
\def\LG{L_{\emph{G}}}

\def\snJ{\textrm{sn}}
\def\cnJ{\textrm{cn}}
\def\dnJ{\textrm{dn}}

\def\dblone{\hbox{$1\hskip -1.2pt\vrule depth 0pt height 1.6ex width 0.7pt \vrule depth 0pt height 0.3pt width 0.12em$}}

\numberwithin{equation}{section}

\begin{document}

\allowdisplaybreaks

\renewcommand{\thefootnote}{$\star$}

\renewcommand{\PaperNumber}{001}

\FirstPageHeading

\ShortArticleName{B\"acklund Transformations  for the Kirchhof\/f Top}

\ArticleName{B\"acklund Transformations  for the Kirchhof\/f Top\footnote{This
paper is a contribution to the Proceedings of the Conference ``Integrable Systems and Geomet\-ry'' (August 12--17, 2010, Pondicherry University, Puducherry, India). The full collection is available at \href{http://www.emis.de/journals/SIGMA/ISG2010.html}{http://www.emis.de/journals/SIGMA/ISG2010.html}}}

\Author{Orlando RAGNISCO and Federico ZULLO}

\AuthorNameForHeading{O.~Ragnisco and F.~Zullo}

\Address{Dipartimento di Fisica Universit\'a Roma Tre and Istituto Nazionale di Fisica Nucleare,\\ Sezione di Roma, I-00146 Roma, Italy}

\Email{\href{mailto:ragnisco@fis.uniroma3.it}{ragnisco@fis.uniroma3.it}, \href{mailto:zullo@fis.uniroma3.it}{zullo@fis.uniroma3.it}}

\URLaddress{\url{http://webusers.fis.uniroma3/~ragnisco/}}

\ArticleDates{Received July 20, 2010, in f\/inal form December 14, 2010;  Published online January 03, 2011}

\Abstract{We construct B\"acklund transformations (BTs) for the Kirchhof\/f top by ta\-king advantage of the common algebraic Poisson structure between this system and the $sl (2)$ trigonometric Gaudin model. Our BTs are integrable maps providing an exact time-discretization of the system, inasmuch as they preserve both its Poisson structure and its invariants. Moreover, in some special cases we are able to show  that these maps  can be explicitly integrated in terms of  the initial conditions and of the ``iteration time''~$n$. Encouraged by these partial results we make the conjecture that the maps are interpolated by a~specif\/ic one-parameter family of hamiltonian f\/lows, and present  the corresponding solution. We enclose a few pictures where the orbits of the continuous  and of the discrete f\/low are depicted.}

\Keywords{Kirchhof\/f equations; B\"acklund transformations;  integrable maps; Lax representation}

\Classification{37J35; 70H06; 70H15}

\renewcommand{\thefootnote}{\arabic{footnote}}
\setcounter{footnote}{0}

\section{Introduction} \label{sec1}
The Kirchhof\/f top is an integrable case of the Kirchhof\/f equations
\cite{Kirchhoff} describing the motion of a~solid in an inf\/inite incompressible
f\/luid.
In general the total kinetic energy of the system
\emph{solid}~$+$ \emph{fluid} is given by a quadratic expression both  in the
translational velocity $\mathbf{u}$ of the rigid body relative to a f\/ixed
frame and
in  its angular velocity $\boldsymbol{\omega}$ \cite{Milne}.
If the solid has three perpendicular planes of symmetry and is one of
revolution too, say
around the $z$ axis, or is a~right prism whose section is any regular
polygon, then  the total kinetic energy reduces to the simple diagonal form~\cite{Lamb}:
\begin{gather}\label{KEC}
T=\frac{1}{2}\big(A_{1}(u_{1}^{2}+u_{2}^{2})+A_{3}u_{3}^{2}\big)+\frac{1}{2}\big(B_{1}(\omega_{1}^{2}+\omega_{2}^{2})+B_{3}\omega_{3}^{2}\big),
\end{gather}
where the quantities $A_{1}$, $A_{3}$, $B_{1}$, $B_{3}$ are constants depending on
the particular shape of the solid.
The total impulse~$\mathbf{p}$ and angular momentum~$\mathbf{J}$ of the
system, i.e.\ the sum of the
impulse and angular momentum of the solid and those applied by the solid to
the boundary of the f\/luid in contact with it, are given by~\cite{Milne}:
\begin{gather*}
p_{i}=\frac{\partial T}{\partial u_{i}}, \qquad J_{i}=\frac{\partial T}{\partial \omega_{i}}.
\end{gather*}
By an Hamiltonian point of view,   impulse and angular momentum must obey  the Lie--Pois\-son~$e(3)$ algebra given
by the following Poisson brackets:
\begin{gather}\label{Poisson}
\{J_{i},J_{j}\}=\epsilon_{ijk}J_{k}, \qquad
\{J_{i},p_{j}\}=\epsilon_{ijk}p_{k}, \qquad \{p_{i},p_{j}\}=0.
\end{gather}
where $i$, $j$, $k$ belong to the set $\{1,2,3\}$. These brackets have two Casimirs:
\begin{gather}\label{Casimir}
\sum_{i=1}^{3}p_{i}J_{i}\doteq C_{1}, \qquad \sum_{i=1}^{3}p_{i}^{2}\doteq 2C_{2}.
\end{gather}
Rewriting the kinetic energy (\ref{KEC}) in terms of the $p_{i}$'s and
$J_{i}$'s, one has two commuting integrals of motion for the Kirchhof\/f top:
\begin{gather*}
T=\frac{1}{2}\left(\frac{p_{1}^2+p_{2}^2}{A_{1}}+\frac{p_{3}^2}{A_{3}}
\right)+\frac{1}{2}\left(\frac{J_{1}^2+J_{2}^2}{B_{1}}+\frac{J_{3}^2}{B_{3}}
\right), \qquad \textrm{and}\qquad J_{3}, \qquad \{T,J_{3}\}=0.
\end{gather*}
The f\/low with respect to the Hamiltonian $T$ is given by the expressions
\begin{gather}\label{eqmotion}
\dot{\mathbf{p}}=\{T,\mathbf{p}\},\qquad \dot{\mathbf{J}}=\{T,\mathbf{J}\}.
\end{gather}

\section{The Kirchhof\/f top by a contraction\\ of trigonometric Gaudin model}\label{sec3}

In this section we show how to obtain the Lax matrix for the Kirchhof\/f top,  in all the cases when the relation $B_{1}^{-1}=A_{3}^{-1}-A_{1}^{-1}$ holds, by a procedure of \textit{pole-coalescence} on the Lax
matrix of the two-site trigonometric Gaudin model~\cite{Gaudin}. The main results are derived
in \cite{MPR,PR}. To this aim, let us brief\/ly review some relevant features of the
trigonometric Gaudin model. In the two-spin case  the Lax matrix reads:
\begin{gather}\label{eq:laxG}
\LG(\lm) =  \left( \begin{array}{cc} \AG(\lm) & \BG(\lm)\\ \CG(\lm)&-\AG(\lm)\end{array}
\right),
\\
\label{eq:ABC}
\AG(\lm)=\sum_{j=1}^{2}\!\cot(\lm-\lm_{j})s^{3}_{j},\!\! \qquad
\BG(\lm)=\sum_{j=1}^{2}\!\frac{s^{-}_{j}}{\sin(\lm-\lm_{j})},\!\!\qquad \CG(\lm)=\sum_{j=1}^{2}\!\frac{s^{+}_{j}}{\sin(\lm-\lm_{j})}.\!\!\!
\end{gather}
In (\ref{eq:laxG}) and (\ref{eq:ABC}) $\lm \in \mathbb{C}$ is the spectral
parameter, $\lm_{j}$ are the arbitrary parameters of the Gaudin model,
 while $\big(s^{+}_{j},s^{-}_{j},s^{3}_{j}\big)$,  $j=1, \ldots, 2$,
are the spin variables of the system obeying to $\oplus^{2} sl(2)$ algebra,
i.e.
\begin{gather} \label{poisS}
\big\{s^{3}_{j},s^{\pm}_{k}\big\}=\mp i\delta_{jk}s^{\pm}_{k}, \qquad
\big\{s^{+}_{j},s^{-}_{k}\big\}=-2i\delta_{jk}s^{3}_{k}.
\end{gather}
In terms of the $r$-matrix formalism, the Lax matrix (\ref{eq:laxG}) satisf\/ies  the \emph{linear} $r$-matrix Poisson algebra:
\begin{gather*} 
\big\{ \LG(\lm), \LG(\mu)\big\}=\big[r_{t}(\lm-\mu), \LG(\lm)\otimes I + I\otimes \LG(\mu) \big],
\end{gather*}
where $r_{t}(\lm)$ stands for the trigonometric $r$ matrix \cite{FT}:
\begin{gather*}
r_{t}(\lm) = \frac{i}{\sin(\lm)}\left(\begin{array}{cccc} \cos(\lm)& 0 & 0 &0 \\0 & 0 & 1 & 0\\
0 & 1 & 0 & 0\\
0 & 0 & 0 & \cos(\lm)
\end{array}\right).
\end{gather*}
The determinant of the Lax matrix (\ref{eq:laxG}) is a generating function of
the integrals of motion. In fact we can write:
\begin{gather*}
-\det(\LG(\lm))=\frac{C_{1\emph{G}}}{\sin(\lm-\lm_{1})^{2}}+\frac{C_{2\emph{G}}}
{\sin(\lm-\lm_{2})^{2}}+\frac{H_{\emph{G}} \sin(\lm_{1}-\lm_{2})}{\sin(\lm-\lm_{1})\sin(\lm-\lm_{2})}-H_{0}^{2},
\end{gather*}
where $C_{1\emph{G}}$ and $C_{2\emph{G}}$ are the Casimirs of the algebra
(\ref{poisS}) given by $C_{i\emph{G}}=(s_{i}^{3})^{2}+s_{i}^{+}s_{i}^{-}$,
while the two involutive integrals of motion $H_{\emph{G}}$ and $H_{0}$ are:
\begin{gather*} 
H_{G}=\frac{2\cos(\lm_{1}-\lm_{2})s_{1}^{3}s_{2}^{3}+
  s_{1}^{+}s_{2}^{-}+ s_{1}^{-}s_{2}^{+}}{\sin(\lm_{1}-\lm_{2})}, \qquad
H_{0}=s_{1}^{3}+s_{2}^{3} \doteq J_{G}^{3},\qquad \{H_{G},H_{0}\}=0.\!\!\!
\end{gather*}
To get the Kirchhof\/f top  we perform the pole-coalescence by introducing  the contraction parameter
$\epsilon$ and take in the Lax matrix (\ref{eq:laxG}) $\lambda_{1}\to\epsilon
\lambda_{1}$ and $\lambda_{2}\to\epsilon\lambda_{2}$.
The Lax matrix for the Kirchhof\/f top is recovered by setting:
 (the notation is
$v_{i}^{\pm}=v_{i}^{1}\pm iv_{i}^{2}$,
$\mathbf{v}_{i}=(v_{i}^{1},v_{i}^{2},v_{i}^{3})$ for any vector set~$\mathbf{v}_{i}$):
\begin{gather}\label{Jp}
\mathbf{J}\doteq \mathbf{s}_{1}+\mathbf{s}_{2}, \qquad \mathbf{p}\doteq
\epsilon (\lm_{1}\mathbf{s}_{1}+\lm_{2}\mathbf{s}_{2})
\end{gather}
and letting $\epsilon \to 0$ in (\ref{eq:laxG}) {\it{after}} this
identif\/ication.  By using (\ref{poisS}), it is readily seen that the
variables $\mathbf{J}$ and $\mathbf{p}$ (\ref{Jp}), obey the Lie--Poisson algebra $e(3)$ (\ref{Poisson}).
Finally,  the Lax matrix for the Kirchhof\/f top reads:
\begin{gather}\label{eq:lax}
L(\lm) =  \left( \begin{array}{cc} A(\lm) & B(\lm)\\ C(\lm)&-A(\lm)\end{array}
\right)=\left(\begin{array}{cc}
  \cot(\lm)J^{3}+\dfrac{p^{3}}{\sin(\lm)^{2}}&\dfrac{J^{-}}{\sin(\lm)}+\dfrac{\cot(\lm)p^{-}}{\sin(\lm)}\vspace{1mm}\\
\dfrac{J^{+}}{\sin(\lm)}+\dfrac{\cot(\lm)p^{+}}{\sin(\lm)}&-(\cot(\lm)J^{3}+\dfrac{p^{3}}{\sin(\lm)^{2}})\end{array}\right).
\end{gather}
Again, its  determinant  is the generating function of the integrals of
motions. Indeed we have:
\begin{gather}\label{genfun}
-\det(L(\lm))=\frac{2H_{1}}{\sin(\lm)^{2}}+2H_{0}\cot(\lm)^2+2C_{2}\frac{\cot(\lm)^{2}}{\sin(\lm)^{2}}+2C_{1}\frac{\cot(\lm)}{\sin(\lm)^{2}},
\end{gather}
where $C_{1}$ and $C_{2}$ are the Casimirs (\ref{Casimir}), while $H_{0}$ and
$H_{1}$ are the two commuting integrals given~by:
\begin{gather}\label{integrals}
H_{1}=\frac{1}{2}\big(J_{1}^{2}+J_{2}^{2}+p_{3}^{2}\big), \qquad 2H_{0}=J_{3}^{2},
\qquad \{H_{1},H_{0}\}=0.
\end{gather}
In all cases where $B_{1}^{-1}=A_{3}^{-1}-A_{1}^{-1}$, the
total kinetic energy (\ref{KEC}) can be rewritten in terms of the quantities (\ref{Casimir}), (\ref{integrals}):
\begin{gather}\label{equiv}
T=\frac{C_{2}}{A_{1}}+\frac{H_{0}}{B_{3}}+\frac{H_{1}}{B_{1}}.
\end{gather}

\section{B\"acklund transformations}

In this section we construct a two parameter family of B\"acklund
Transformations def\/ining symplectic, integrable and explicit maps that,  as
we will see, provide an exact  time-discretisation of our model. The
approach follows that given for example in~\cite{KS} and take advantage of the
results derived in~\cite{RZ} where the B\"acklund transformations (BT) for the $N$-site
trigonometric Gaudin magnet have been constructed. In fact, since  the $r$-matrix structure survives the pole-coalescence and contraction procedures, the
ans\"atze for the dressing matrix $D(\lambda)$ linking, by a~similarity
transformation, the \textit{old} Lax matrix $L(\lambda)$ to the \textit{new} Lax matrix
$\tilde{L}(\lambda)$ are the same as for the trigonometric Gaudin.
Thus, according to the procedure followed in  \cite{RZ},  we write:
\begin{gather}\label{simil}
\tilde{L}(\lambda)D(\lambda)=D(\lambda)L(\lambda),
\end{gather}
where $\tilde{L}$ has the same $\lambda$ dependence as in (\ref{eq:lax}) but
is written in terms of the updated variables ($\tilde{J}^{3}, \tilde{J}^{\pm},
\tilde{p}^{3}, \tilde{p}^{\pm}$). The matrix $D(\lambda)$ reads \cite{RZ}
\begin{gather} \label{eq:Darboux}
D(\lambda)=\left(\begin{array}{cc} \sin(\lambda-\lambda_{0}-\mu)+PQ \cos(\lambda-\lambda_{0}) &
  P \cos(\mu)\\
Q \sin(2\mu)-PQ^{2}\cos(\mu) & \sin(\lambda-\lambda_{0}+\mu)-PQ \cos(\lambda-\lambda_{0})
\end{array}\right).
\end{gather}
In (\ref{eq:Darboux})  $\lambda_{0}$ and $\mu$ are arbitrary constants and $P$ and $Q$ are, up to now, indeterminate dynamical variables.

We remark that in the fundamental paper by V.~Kuznetsov and P.~Vanhaecke~\cite{KV}, an extensive study of BTs in the $2\times 2$ case has been performed. In fact a reformulation of our similarity transformation (\ref{simil}) having a polynomial dependence on the spectral parameter can be derived starting from their results.

Our aim is now to f\/ind an expression for $P$ and $Q$ in terms of only one set of
dynamical variables, say the old ones, so that (\ref{simil}) yields
 the explicit map between the two sets of variables. To achieve this goal, we use
the so-called spectrality property (see for example \cite{KS}).

Note that the
determinant of $D(\lambda)$ is proportional to
 $\sin(\lambda-\lambda_{0}-\mu)\sin(\lambda-\lambda_{0}+\mu)$, so, modulo $2\pi$,  it has two zeros,  $\lambda_+=\lambda_{0}+\mu$ and
$\lambda_-=\lambda_{0}-\mu$.  $D(\lambda_\pm)$
are clearly  rank one matrices, having  one dimensional kernels, say,  $|K_{\pm}\rangle$. The key point is that these kernels are eigenvectors of the Lax matrix. Indeed from (\ref{simil}) it follows:
\begin{gather}\label{KpKm}
L(\lambda_\pm)|K_{\pm}\rangle=\gamma_{\pm}|K_{\pm}\rangle,
\end{gather}
where the two eigenvalues are given by:
\begin{gather*}
\gamma_{\pm}^{2}=A^{2}(\lambda)+B(\lambda)C(\lambda)\big|_{\lambda=\lambda_\pm}
\end{gather*}
and $A(\lambda)$, $B(\lambda)$ and $C(\lambda)$ are def\/ined in (\ref{eq:lax}). The equation~(\ref{KpKm}) gives the relations between~$P$,~$Q$ and the old dynamical
variables. In fact,  the two kernels are given by:
\begin{gather*}
|K_{+}\rangle=\left(\begin{array}{c}1\\-Q\end{array}\right), \qquad |K_{-}\rangle=\left(\begin{array}{c}P\\2\,\sin(\mu)-PQ\end{array}\right)
\end{gather*}
and then readily follow the expressions for $Q$ and $P$:
\begin{gather*}
Q=Q(\lambda_+), \qquad
\frac{1}{P}=\frac{Q(\lambda_+)-Q(\lambda_-)}{2\,\sin(\mu)},\qquad Q(\lambda_{\pm})=\frac{A(\lambda_{\pm})\mp\gm(\lambda_{\pm})}{B(\lambda_{\pm})}.
\end{gather*}
Taking the residue of  (\ref{simil})  at the pole in $\lambda=0$ and its value at
$\lambda=\frac{\pi}{2}$ we obtain the explicit maps as below:
\begin{gather}
 \tilde{p}^{-}=\frac{1}{\Delta\sin(\lambda_+)\sin(\lambda_-)}\big(a_{+}^{2}p^{-}-P^{2}\cos(\mu)^{2}p^{+}+2P\cos(\mu)a_{+}p^{3}\big),\nonumber\\
 \tilde{p}^{+}=\frac{1}{\Delta\sin(\lambda_+)\sin(\lambda_-)}\big(a_{-}^{2}p^{+}-Q^{2}\cos(\mu)^{2} c^{2}p^{-}-2Q\cos(\mu) c a_{-}p^{3}\big),\nonumber\\
 \tilde{p}^{3}=\frac{1}{\Delta\sin(\lambda_+)\sin(\lambda_-)}\big(2a_{+}a_{-}p^{3}-P\cos(\mu)a_{-}p^{+}
+Q\cos(\mu)\,c\,a_{+}p^{-}\big)-p^{3},\nonumber\\
 \tilde{J}^{-}=\frac{1}{\Delta\cos(\lambda_+)\cos(\lambda_-)}\big(b_{+}^{2}J^{-}-P^{2}\cos(\mu)^{2}J^{+}
-2P\cos(\mu)b_{+}p^{3}\big),\nonumber\\
 \tilde{J}^{+}=\frac{1}{\Delta\cos(\lambda_+)\cos(\lambda_-)}\big(b_{-}^{2}J^{+}-Q^{2}\cos(\mu)^{2} c^{2}J^{-}
+2Q\cos(\mu)\,c\,b_{-}p^{3}\big),\nonumber\\
 \tilde{J}^{3}=J^{3},\label{maps}
\end{gather}
where
\begin{gather*}
 a_{\pm}\doteq\sin(\lambda_{\pm})\mp PQ\cos(\lambda_{0}),\qquad b_{\pm}\doteq\cos(\lambda_{\pm})\pm PQ\cos(\lambda_{0}), \\
 \Delta\doteq 1-2PQ\sin(\mu)+P^{2}Q^{2},\qquad c\doteq 2\sin(\mu)-PQ.
 \end{gather*}
Thus the maps depend on two B\"acklund
parameters, $\lambda_{0}$ and $\mu$ (or $\lambda_+$ and $\lambda_-$): in
the next section we will show that, provided $\lambda_{0}\in\mathbb{R}$ and
$\mu \in i\mathbb{R}$, this two-point transformation is actually a time discretization of a one parameter family of
continuous f\/lows having the same integrals of motion~(\ref{Casimir}),
(\ref{integrals}) as the continuous dynamical system ruled by the
physical Hamiltonian (\ref{equiv}). With the above constraints on the parameters, the BTs become physical, mapping  real variables into  real variables.
Furthermore these transformations are symplectic. In fact, as the $r$-matrix structure underlying the Kirchhof\/f top is the same as that of the ancestor trigonometric Gaudin magnet, the simplecticity of the transformations (\ref{maps}) is guaranteed along the lines given in~\cite{RZ}.

Next,  we will formulate  the conjecture  that, provided $\lambda_{0}\in\mathbb{R}$ and
$\mu \in i\mathbb{R}$, this two-point transformation is not only a time discretization of a one parameter family of continuous f\/lows equipped with the same integrals of motion (\ref{Casimir}),
(\ref{integrals}), but it has also the same orbits as the continuous dynamical system ruled by the
physical Hamiltonian (\ref{equiv}). The above conjecture will be verif\/ied to hold in a couple of special cases, where the explicit solution of the recurrences def\/ined by the maps (\ref{maps}) will be derived, and shown to be interpolated by the solution to the evolution equations for the continuous Kirchhof\/f top. On  one hand this conf\/irms the Kuznetsov--Sklyanin intuition that
B\"acklund transformations can be used as a tool for separation of variables (see also~\cite{KV}), on the  other hand, in light of~\cite{KV}, these results appear quite natural if one thinks that the B\"acklund transformations are translations on the invariant tori, translations that corresponds to some addition formulas  for families of hyperelliptic functions.

\section{Continuum limit and discrete dynamics}
As shown in \cite{RZ},  to ensure ``reality'' of the maps  (\ref{maps}),  one has to require  the Darboux
matrix~$D$  to be  a unitary matrix (possibly up to an irrelevant scalar factor); this holds true if\/f
 $\lambda_\pm$  are mutually  complex conjugate,
i.e.\ if\/f  $\lambda_{0}$ is real and  $\mu$ is pure imaginary. So we set:
\begin{gather*}
\lambda_{+}=\lambda_{0}+i\frac{\epsilon}{2}, \qquad \lambda_{-}=\lambda_{0}-i\frac{\epsilon}{2}.
\end{gather*}
In the limit $\epsilon \to 0$ the relations
(\ref{maps}) go into the identity map. Indeed $\epsilon$
plays the role of time step for the one parameter ($\lambda_{0}$) discrete dynamics def\/ined by the
B\"acklund transformations. By following~\cite{RZ}, in order to identify the continuous limit of this
discrete dynamics we take the Taylor expansion of the dressing matrix at order
$\epsilon$, obtaining:
\begin{gather*}
D(\lambda)=\sin(\lambda-\lambda_{0})\dblone
-\frac{i\epsilon }{2\gamma(\lambda_{0})}\left(\begin{array}{cc}
A(\lambda_{0})\cos(\lambda-\lambda_{0}) & B(\lambda_{0})\\
C(\lambda_{0})& -A(\lambda_{0})\cos(\lambda-\lambda_{0})
\end{array}\right)+O\big(\epsilon^{2}\big),
\end{gather*}
where the functions $A(\lambda)$, $B(\lambda)$ and $C(\lambda)$ are given by
(\ref{eq:lax}), and $\gamma(\lambda)^{2}=A(\lambda)^{2}+B(\lambda)C(\lambda)$.
By inserting this expression in the equation (\ref{simil}) we arrive at the Lax pair for
the continuous f\/low:
\begin{gather}\label{eq:motion}
\dot{L}(\lm)= [L(\lambda),M(\lambda,\lambda_{0})],
\end{gather}
where the ``time derivative'' is def\/ined as $\dot{L}=\lim\limits_{\epsilon\rightarrow 0}\frac{\tilde{L}-L}{\epsilon}$.

The matrix $M(\lambda,\lambda_{0})$ takes the explicit form:
\begin{gather*}
M(\lambda,\lambda_{0})=\frac{i}{2\gamma(\lambda_{0})}\left(\begin{array}{cc}
A(\lambda_{0})\cot(\lambda-\lambda_{0}) & \dfrac{B(\lambda_{0})}{\sin(\lambda-\lambda_{0})}\\
\dfrac{C(\lambda_{0})}{\sin(\lambda-\lambda{0})}& -A(\lambda_{0})\cot(\lambda-\lambda_{0})
\end{array}\right).
\end{gather*}
In Hamiltonian terms,  the system (\ref{eq:motion})
reads:
\begin{gather}\label{eqLp}
\dot{L}_{ij}(\lambda)=\{\gamma(\lambda_{0}),L_{ij}(\lambda)\}, \qquad i,j \in (1,2),
\end{gather}
entailing that  the variables $\mathbf{p}$ and $\mathbf{J}$ of the continuous f\/low obey the evolution equations:
\begin{gather}\label{eqmotiond}
\dot{\mathbf{p}}=\{\gamma(\lm_0),\mathbf{p}\},\qquad \dot{\mathbf{J}}=\{\gm(\lm_0),\mathbf{J}\}.
\end{gather}
It is clear that the dynamical system given by (\ref{eqmotiond}) possesses  the
integrals (\ref{integrals}), because of (\ref{genfun}). Moreover we have some  evidences, that will be reported in the following, that the continuous and the discrete system share the same orbits too.

First of all we note that the direction of the continuous f\/low that obtains in the continuum limit from the discrete dynamics def\/ined by the B\"acklund transformations (\ref{maps}), and that of the Kirchhof\/f top (\ref{eqmotion}) with the kinetic energy $T$ given by (\ref{equiv}), can be made parallel.  In fact the shape of the
orbits are unchanged if one takes an arbitrary $C^{1}$ function of the Hamiltonian
$\gamma(\lambda_{0})$ as a new Hamiltonian in (\ref{eqLp}),
since  this  operation amounts just  to a  time rescaling
(for every f\/ixed orbit $\gamma(\lambda_{0})$ is constant). Accordingly, we take as Hamiltonian function
 $\frac{w\gamma(\lambda_{0})^{2}}{2}$, where $w$ is,
so far, an arbitrary constant. The expression (\ref{genfun}) allows to write
the explicit equations of motion for a generic function of the dynamical
variables $\mathcal{F}(\mathbf{p},\mathbf{J})$:
\begin{gather*}
\dot{\mathcal{F}}(\mathbf{p},\mathbf{J})=\left\{\frac{w\gamma(\lambda_{0})^{2}}{2},\mathcal{F}(\mathbf{p},\mathbf{J})\right\}
=\left\{w\frac{H_{1}}{\sin(\lambda_{0})^{2}}+w\frac{H_{0}\cos(\lambda_{0})^{2}}{\sin(\lambda_{0})^{2}},\mathcal{F}(\mathbf{p},\mathbf{J})\right\}.
\end{gather*}
This has to be compared with with the equations of motion for the physical
Hamiltonian (\ref{equiv}):
\begin{gather*}
\dot{\mathcal{F}}(\mathbf{p},\mathbf{J})=\left\{\frac{H_{1}}{B_{1}}+\frac{H_{0}}{B_{3}},\mathcal{F}(\mathbf{p},\mathbf{J})\right\}.
\end{gather*}
The two expressions coincide by identifying:
\begin{gather}\label{sinlm}
w=\frac{1}{B_{1}}-\frac{1}{B_{3}},\qquad \sin(\lambda_{0})^{2}=\frac{B_{3}-B_{1}}{B_{3}}.
\end{gather}
In other words, the physical f\/low is the continuum limit of the  discretized one. Now we make the following

\medskip

\noindent
\textbf{Conjecture.} \emph{For any fixed $\lm_0$, there exist a re-parametrization of $\epsilon$, $\epsilon \to T_{\lm_0}$, possibly depending  by the integrals and the Casimir functions, such that $T=\epsilon +O(\epsilon^2)$, and at all order in $T$ the continuous orbits of the physical flow interpolate the discrete orbits defined by the B\"acklund transformations,  provided that $\lm_0$ is chosen according to \eqref{sinlm}.}

\medskip

This is equivalent to say that, for any f\/ixed $\lm_0$, via the above reparameterization,  B\"acklund transformations form  a one parameter ($T$) group of transformations, obeying the linear composition law $\textrm{BT}_{T_1}\circ \textrm{BT}_{T_2}=\textrm{BT}_{T_1+T_2}$,  ``$\circ$'' being the composition. Note also that, \emph{if the conjecture is true},  then, since at f\/irst order in $\epsilon$ (and therefore in $T$) the f\/low is ruled by the Hamiltonian~$\gm(\lm_0)$, one has:
\[
\tl{x}^n=e^{nT\{\gm(\lm_0), \cdot\}}x=x+nT\{\gm(\lm_0),x\}+\frac{n^2T^2}{2}\{\gamma(\lambda_0),\{\gamma(\lambda_0),x\}\}+\cdots,
\]
where $\tl{x}^n$ means the $n$-th iteration of the B\"acklund transformations with the same parameter~$T$.

\begin{figure}[t]
\centering
\includegraphics[scale=0.6]{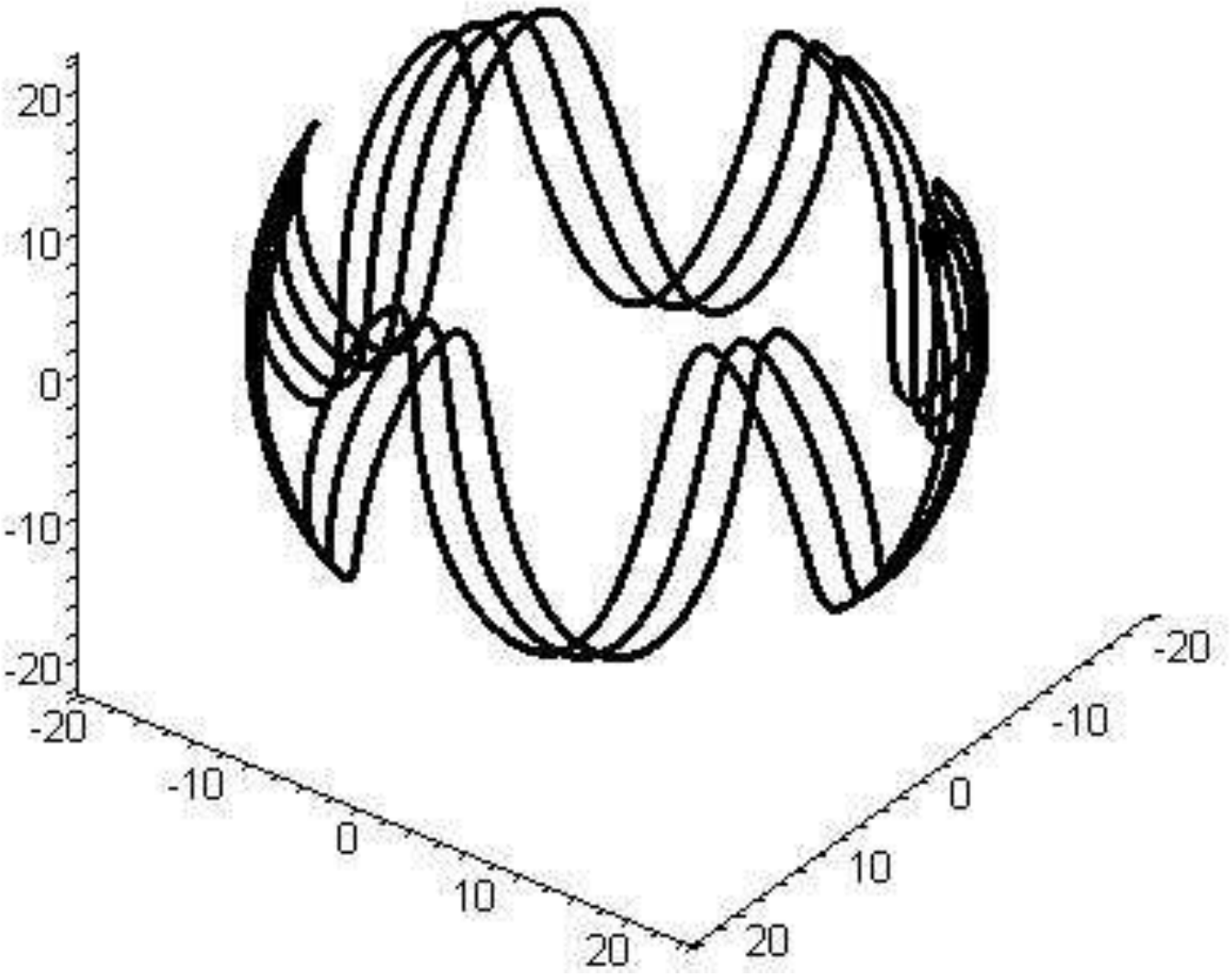}
\caption{initial conditions: $p^{1}(0)=15$, $p^{2}(0)=-12.13$, $p^{3}(0)=-10$,
$J^{1}(0)=1$, $J^{2}(0)=-4$, $J^{3}(0)=3$. Moments of inertia: $B_{1}=1$,
  $B_{3}=\sec(0.1)^{2}$.}
\label{fig:2}
\end{figure}

\begin{figure} [t]
\centering
\includegraphics[scale=0.6]{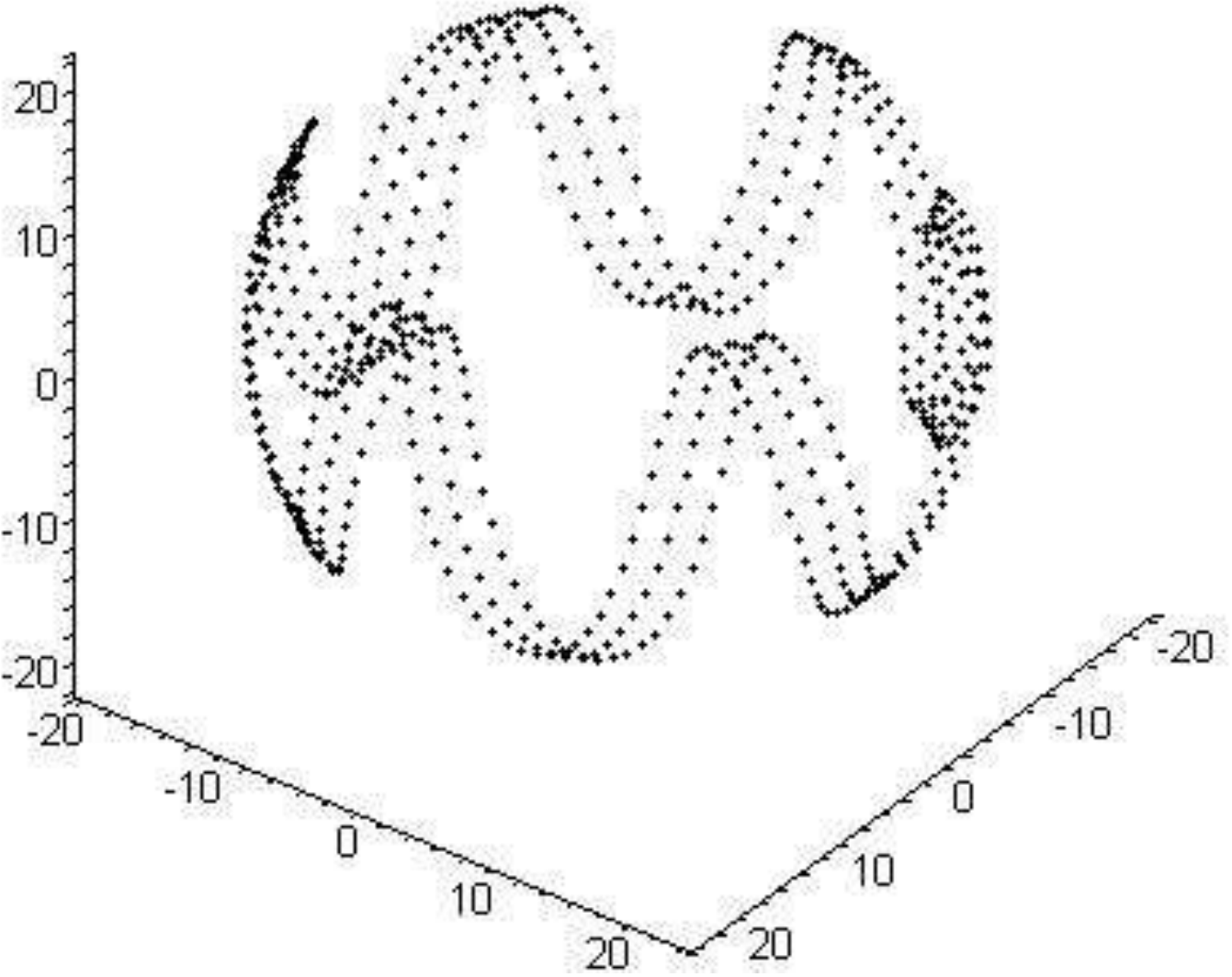}
\caption{input parameters: $p^{1}(0)=15$, $p^{2}(0)=-12.13$, $p^{3}(0)=-10$,
$J^{1}(0)=1$, $J^{2}(0)=-4$, $J^{3}(0)=3$, $\lambda_{0}=0.1$, $\epsilon=0.1$.}
\label{fig:3}
\end{figure}

In the Figs.~\ref{fig:2} and~\ref{fig:3}  we report respectively an example
of the orbit for the variables $(p^{1}(t)$, $p^{2}(t),p^{3}(t))$ for the continuous f\/low ruled by
the Hamiltonian (\ref{equiv}) as given in Appendix~\ref{Appendix}  and of the corresponding discrete
f\/low obtained by iterating the B\"acklund transformations. The initial conditions are the same and the value of $\lambda_0$ has been chosen so to make the continuous limit of the discrete dynamics parallel to the continuous f\/low of the Kirchhof\/f top. They overlap exactly. In the next section, assuming the conjecture to hold true,  we will show a way  to f\/ind the parameter~$T$.
There, we will give as well analytic results in two particular cases, where the continuous f\/low is periodic, and not just quasiperiodic. Clearly, these non generic  examples cannot be invoked to support  our conjecture: however,  we decided to include them in the  paper inasmuch as they provide an explicit link between discrete and continuous  dynamics.

\subsection{Integrating the B\"acklund: special examples}

Let us assume to have a smooth transformation, that we indicate with $\tilde{x}=f(x,\epsilon)$,
where the parameter $\epsilon$ plays the role of the time step, such that
$f(x,0)=x$. By $\tilde{x}^{n}$ we denote the $n$-th iteration of the
map, so that $\tilde{x}^{0}=x$, $\tilde{x}^{1}=f(x,\epsilon)$,
$\tilde{x}^{2}=f(f(x,\epsilon),\epsilon)$ and so on. Solving the B\"acklund map amounts
to f\/ind $\tilde{x}^{n}$ as a function of $x$, $n$ and $\epsilon$.
Now we will show that, under given assumptions, there is indeed a positive answer  to
this question. We will follow a simple argument, well known in group
theory~\cite{Hamermesh}.

Suppose to do a transformation from $x$ to $\tilde{x}^{1}$ with parameter
$\epsilon_{1}$ and then another one from~$\tilde{x}^{1}$ to~$\tilde{x}^{2}$ with
parameter~$\epsilon_{2}$. Suppose also that there exist a parameter~$\epsilon_3$ linking directly~$x$ to~$\tilde{x}^{2}$. As the B\"acklund are
smooth, varying continuously~$\epsilon_{1}$ or~$\epsilon_{2}$ corresponds to a
continuous variation in $\epsilon_{3}$: the B\"acklund
transformations def\/ine $\epsilon_{3}$ as a continuous function of~$\epsilon_{1}$ and
$\epsilon_{2}$, say $\epsilon_{3}=\chi(\epsilon_{1},\epsilon_{2})$. Now consider inf\/initesimal transformations:
a small change in the parameter~$\epsilon$ take the point $\tilde{x}^{1}$ to a near
point $\tilde{x}^{1}+d\tilde{x}^{1}$:
\begin{gather*}
\tilde{x}^{1}+d\tilde{x}^{1}=f(x,\epsilon+d\epsilon).
\end{gather*}
But we can arrive at the same point by starting from $\tilde{x}^{1}$ and
acting on it with a transformation near the identity, say with the small
parameter $\delta\epsilon$:
\begin{gather}\label{xpdx}
\tilde{x}^{1}+d\tilde{x}^{1}=f(\tilde{x}^{1},\delta\epsilon).
\end{gather}
The relation between the parameters now reads:
\begin{gather*}
\epsilon+d\epsilon=\chi(\epsilon,\delta\epsilon).
\end{gather*}
Obviously $\chi(\epsilon,0)=\epsilon$, so:
\begin{gather}\label{dmu}
d\epsilon=\left.\frac{\partial \chi}{\partial
  \delta\epsilon}\right|_{\delta\epsilon=0}\delta\epsilon\doteq \tau(\epsilon)\delta\epsilon.
\end{gather}
The relation (\ref{xpdx}) tells us that:
\begin{gather*}
d\tilde{x}^{1}=\left.\frac{\partial f(\tilde{x}^{1},\delta\epsilon)}{\partial
  \delta\epsilon}\right|_{\delta\epsilon=0}\delta\epsilon\doteq \zeta(\tilde{x}^{1})\delta\epsilon.
\end{gather*}
The last expression together with (\ref{dmu}) gives:
\begin{gather}\label{parameterT}
\int_{x}^{\tilde{x}^{1}}\frac{dy}{\zeta(y)}=\int_{0}^{\epsilon}\frac{d\lambda}{\tau(\lambda)}\doteq T.
\end{gather}
This means that there exists a function, say $V$, such that:
\begin{gather*}
V(\tilde{x}^{1})=V(x)+T \quad \Longrightarrow \quad V(\tilde{x}^{n})=V(x)+nT.
\end{gather*}
Formally we can write this expression as $\tilde{x}^{n}=V^{-1}(V(x)+nT)$. However,
for $n=1$ we must have $\tilde{x}^{1}=f(x,\epsilon(T))$, yielding
$\tilde{x}^{n}=f(x,\epsilon(nT))$. The continuous f\/low discretized is simply given by
$x(t)=f(x,\epsilon(t))$ where $x$ is the initial condition ($x(t=0)=x$).

In the following we will present  two particular cases, both corresponding to periodic f\/lows,  where the
 B\"acklund
transformations can be explicitly integrated.

\begin{example}
Consider the invariant submanifold $\mathbf{p}=(X,0,Z)$,
$\mathbf{J}=(0, Y, 0)$. Since now $H_{0}=0$, the
freedom to have a parameter $\lambda_{0}$ in (\ref{eqLp}) is just a scaling in time,
so we can freely f\/ix it: by now we pose $\lambda_{0}=\frac{\pi}{2}$. With this
choice the interpolating Hamiltonian f\/low discretized by the maps~(\ref{maps})
is given simply by $\mathcal{H}=\sqrt{Y^{2}+Z^{2}}$. So, as seen at the beginning
of this section, in order to have real
transformations we pose $\lambda_{1}=\frac{\pi}{2}+i\epsilon$ and $\lambda_{2}=\frac{\pi}{2}-i\epsilon$. The B\"acklund
transformation can be now conveniently written in terms of a single function~$R$ of~$\epsilon$,~$X$,~$Y$ and~$Z$:
\begin{subequations}\label{newmaps}
\begin{gather}
\tilde{X}=\frac{4R\sinh(\epsilon)(R^{2}+1)}{(R^{2}-1)^{2}+4\cosh(\epsilon)^{2}R^{2}}Z
+\frac{(R^{2}+1)^{2}-4R^{2}\sinh(\epsilon)^{2}}{(R^{2}-1)^{2}+4R^{2}\cosh(\epsilon)^{2}}X,
\\
\tilde{Y}=\frac{4R\cosh(\epsilon)(R^{2}-1)}{(R^{2}-1)^{2}+4\cosh(\epsilon)^{2}R^{2}}Z
-\frac{(R^{2}-1)^{2}-4R^{2}\cosh(\epsilon)^{2}}{(R^{2}-1)^{2}+4R^{2}\cosh(\epsilon)^{2}}Y,
\\
\tilde{Z}=\frac{(R^{2}+1)^{2}-4R^{2}\sinh(\epsilon)^{2}}{(R^{2}-1)^{2}+4R^{2}\cosh(\epsilon)^{2}}Z
-\frac{4R\sinh(\epsilon)(R^{2}+1)}{(R^{2}-1)^{2}+4\cosh(\epsilon)^{2}R^{2}}X,
\\
R\doteq\frac{Z-\sqrt{(\mathcal{H}^{2}\cosh(\epsilon)^{2}-2C_{2}\sinh(\epsilon)^{2})}}{X\sinh(\epsilon)+Y\cosh(\epsilon)}.\nonumber
\end{gather}
\end{subequations}
Note that the two constants under square  root in the numerator of $R$ are the
Hamiltonian $\mathcal{H}=\sqrt{Y^{2}+Z^{2}}$ and the Casimir function
$C_{2}=\frac{X^{2}+Z^{2}}{2}$. To solve the recurrences (\ref{newmaps}) one has to f\/ind $\epsilon$ as a function of the
parameter $T$ def\/ined in (\ref{parameterT}). To this end we f\/irst note that
$\left.\frac{d\tilde{Z}}{d\epsilon}\right|_{\epsilon=0}=\frac{2XY}{\mathcal{H}}$, so
that by the relations (\ref{parameterT}) we have:
\begin{gather}\label{newparT}
\int^{\tilde{Z}}_{Z}\mathcal{H}\frac{d\tilde{Z}}{2\tilde{X}\tilde{Y}}
=\int_{0}^{\epsilon}\mathcal{H}\frac{1}{2\tilde{X}\tilde{Y}}\frac{d\tilde{Z}}{d\epsilon}d\epsilon
=\int_{0}^{\epsilon}\mathcal{H}\frac{d\epsilon}{\sqrt{\mathcal{H}^{2}\cosh(\epsilon)^{2}-2C_{2}\sinh(\epsilon)^{2}}}=T.
\end{gather}
All that we have to do now is to perform the integral, invert the result
to f\/ind $\epsilon$ as a function of~$T$, then plug the result into (\ref{newmaps})
and replace $T$ by $nT$: this gives the solution to the B\"acklund
recurrences. After some manipulations with the Jacobian elliptic functions we
arrive at the simple result:
\begin{gather*}
\cosh(\epsilon)=\frac{1}{\cnJ\big(T,\frac{\sqrt{2C_{2}}}{\mathcal{H}}\big)}, \qquad
\sinh(\epsilon)=\frac{\snJ\big(T,\frac{\sqrt{2C_{2}}}{\mathcal{H}}\big)}{\cnJ\big(T,\frac{\sqrt{2C_{2}}}{\mathcal{H}}\big)}.
\end{gather*}
With this position we can write down the expressions for $\tilde{X}^{n}$,
$\tilde{Y}^{n}$ and $\tilde{Z}^{n}$:
\begin{gather*}
\tilde{X}^{n}=\frac{4R\snJ(nT)\cnJ(nT)(R^{2}+1)}{(R^{2}-1)^{2}\cnJ(nT)^{2}+4R^{2}}Z
+\frac{(R^{2}+1)^{2}\cnJ(nT)^{2}-4R^{2}\snJ(nT)^{2}}{(R^{2}-1)^{2}\cnJ(nT)^{2}+4R^{2}}X,
\\
\tilde{Y}^{n}=\frac{4R\cnJ(nT)(R^{2}-1)}{(R^{2}-1)^{2}\cnJ(nT)^{2}+4R^{2}}Z-\frac{(R^{2}-1)^{2}\cnJ(nT)^{2}-4R^{2}}{(R^{2}-1)^{2}\cnJ(nT)^{2}+4R^{2}}Y,
\\
\tilde{Z}^{n}=\frac{(R^{2}+1)^{2}\cnJ(nT)^{2}-4R^{2}\snJ(nT)^{2}}{(R^{2}-1)^{2}\cnJ(nT)^{2}+4R^{2}}Z
-\frac{4R\snJ(nT)\cnJ(nT)(R^{2}+1)}{(R^{2}-1)^{2}\cnJ(nT)^{2}+4R^{2}}X,
\\
R=\frac{Z\cnJ(nT)-\sqrt{(\mathcal{H}^{2}-2C_{2}\snJ(nT)^{2})}}{X\snJ(nT)+Y},\nonumber
\end{gather*}
where for brevity we have omitted the elliptic modulus
$\frac{\sqrt{2C_{2}}}{\mathcal{H}}$ in the Jacobian elliptic functions ``sn''
and ``cn''. Note that if
we pose in (\ref{newparT}) $2T=t$, that is
\begin{gather*}
\cosh(\epsilon)=\frac{1}{\cnJ\big(\frac{t}{2},\frac{\sqrt{2C_{2}}}{\mathcal{H}}\big)}, \qquad
\sinh(\epsilon)=\frac{\snJ\big(\frac{t}{2},\frac{\sqrt{2C_{2}}}{\mathcal{H}}\big)}{\cnJ\big(\frac{t}{2},\frac{\sqrt{2C_{2}}}{\mathcal{H}}\big)}
\end{gather*}
in (\ref{newmaps}), then we have the \emph{general solution} of the dynamical system
ruled by the interpolating Hamiltonian f\/low $\mathcal{H}=\sqrt{Z^{2}+Y^{2}}$,
 that is the value that takes the hamiltonian $\gm(\lm_0)$ (\ref{eqmotiond})
  on the invariant submanifold considered in this example for $\lm_0$=$\frac{\pi}{2}$.
The equations of motion are given by $\mathcal{H}\dot{X}=-YZ$,
$\mathcal{H}\dot{Y}=-XZ$, $\mathcal{H}\dot{Z}=XY$.

Obviously this general solution coincide with that found by a direct
integration of the previous equation of motion, i.e.\ with
$Z=\sqrt{2C_{2}}\,\snJ(t+v)$, $X=\sqrt{2C_{2}}\,\cnJ(t+v)$ and
$Y=\mathcal{H}\dnJ(t+v)$, where the elliptic modulus of this functions is
again $\frac{\sqrt{2C_{2}}}{\mathcal{H}}$ and where $v$ is such that $\snJ(v)=\frac{Z}{\sqrt{2C_{2}}}$.
\end{example}

\begin{example}
In the next example we consider the invariant submanifold $\mathbf{p}=(x,y,0)$,
$\mathbf{J}=(0, 0, z)$. Again, in order to have real
transformations we pose $\lambda_{1}=\lambda_0+i\epsilon$ and $\lambda_{2}=\lambda_0-i\epsilon$ with $\lm_0$ and $\epsilon$ real. In terms of $p^{\pm}=x\pm i\,y$, the maps (\ref{maps}) become:
\begin{gather}
\tl{p}^{-}=\frac{z \sin(\lm_0-i\epsilon)-\sqrt{z^2\sin(\lm_0-i\epsilon)^2+p^-p^+}}{z \sin(\lm_0+i\epsilon)-\sqrt{z^2\sin(\lm_0+i\epsilon)^2+p^-p^+}}p^{-},\nonumber\\
\tl{p}^{+}=\frac{z \sin(\lm_0+i\epsilon)-\sqrt{z^2\sin(\lm_0+i\epsilon)^2+p^-p^+}}{z \sin(\lm_0-i\epsilon)-\sqrt{z^2\sin(\lm_0-i\epsilon)^2+p^-p^+}}p^{+},\qquad
\tl{z}=z.\label{harmaps} 
\end{gather}
To f\/ind the relation def\/ining the parameter $T$ (\ref{parameterT}), f\/irst f\/ind the expression of $\left.\frac{d\tilde{p}^-}{d\epsilon}\right|_{\epsilon=0}$:
\[
\left.\frac{d\tilde{p}^-}{d\epsilon}\right|_{\epsilon=0}=\frac{2iz\cos(\lm_0)p^-}{\sqrt{z^2\sin(\lm_0)^2+p^+p^-}},
\] then by using (\ref{parameterT}) one has:
\begin{gather*}
\frac{\sqrt{z^2\sin(\lm_0)^2+p^+p^-}}{2\cos(\lm_0)}\int^{\tilde{p}^-}_{p^-}\frac{d\tilde{p}^-}{iz\tilde{p}^-}=T=
 \frac{\sqrt{z^2\sin(\lm_0)^2+p^+p^-}}{2\cos(\lm_0)} \nonumber\\
 \qquad{}\times\int_{0}^{\epsilon}\left(\frac{\cos(\lm_0+i\epsilon)}{\sqrt{z^2\sin(\lm_0+i\epsilon)^2+p^+p^-}}
 +\frac{\cos(\lm_0-i\epsilon)}{\sqrt{z^2\sin(\lm_0-i\epsilon)^2+p^+p^-}}\right)d \epsilon
\end{gather*}
or more explicitly:
\begin{gather*}
 \frac{2 i \cos(\lm_0)T}{\sqrt{z^2\sin(\lm_0)^2+p^+p^-}}=\textrm{arcsinh}\left(\frac{z}{\sqrt{p^+p^-}}\sin(\lm_0+i \epsilon)\right)\!-\textrm{arcsinh}\left(\frac{z}{\sqrt{p^+p^-}}\sin(\lm_0-i\,\epsilon)\right)\nonumber\\
\hphantom{\frac{2 i \cos(\lm_0)T}{\sqrt{z^2\sin(\lm_0)^2+p^+p^-}}}{}
 =\ln\left(\frac{z \sin(\lm_0-i\epsilon)-\sqrt{z^2\sin(\lm_0-i\epsilon)^2+p^-p^+}}{z \sin(\lm_0+i\epsilon)-\sqrt{z^2\sin(\lm_0+i\epsilon)^2+p^-p^+}}\right).
\end{gather*}
The B\"acklund transformations (\ref{harmaps}) now take the simple form:
 \begin{gather*}
 \tl{p}^-=\exp\left(\frac{2 i \cos(\lm_0)zT}{\sqrt{z^2\sin(\lm_0)^2+p^+p^-}}\right)p^-,\qquad
 \tl{p}^+=\exp\left(-\frac{2 i \cos(\lm_0)zT}{\sqrt{z^2\sin(\lm_0)^2+p^+p^-}}\right)p^+,\\
 \tl{z}=z,\nonumber
\end{gather*}
so that again, as expected, the $n$-th iteration of the maps $(\tl{p}^{-})^n$, $(\tl{p}^+)^n$, $\tl{z}^n$ is found by substituting~$T$ with~$nT$.  By posing $2T=t$ in the previous expressions and returning to the real variables $x=\frac{p^++p^-}{2}$ and $y=\frac{p^+-p^-}{2 i}$, we have the continuous f\/low:
\begin{gather*}
x(t)=x\cos\left(\frac{\cos(\lm_0)z}{\sqrt{z^2\sin(\lm_0)^2+p^+p^-}}t\right)+y\sin\left(\frac{\cos(\lm_0)z}{\sqrt{z^2\sin(\lm_0)^2+p^+p^-}}t\right),\\
y(t)=y\cos\left(\frac{\cos(\lm_0)z}{\sqrt{z^2\sin(\lm_0)^2+p^+p^-}}t\right)-x\sin\left(\frac{\cos(\lm_0)z}{\sqrt{z^2\sin(\lm_0)^2+p^+p^-}}t\right)
 \end{gather*}
corresponding to the general solution of the continuous system ruled by the value that takes the hamiltonian $\gm(\lm_0)$ (\ref{eqmotiond}) on the invariant submanifold $\mathbf{p}=(x,y,0)$, $\mathbf{J}=(0, 0, z)$ considered in this example.
\end{example}

\appendix
\section{Integration of the continuous  model}\label{Appendix}

For the sake of completeness, we report here the solution to the evolution equations def\/ining the Kirchhof\/f top.  Gustav Kirchhof\/f in his ``\emph{Vorlesungen \"uber mathematische Physik}'' \cite{Kirchhoff} deals with
the problem of integrating equations of motion (\ref{eqmotion}). For the
physical variables $(J^{1},J^{2},J^{3})$ and $(p^{1},p^{2},p^{3})$ they are
written in extended form as:
\begin{gather*}
 \dot{p}^{1}(t)=\alpha J^{3}(t)p^{2}(t)-\beta J^{2}(t)p^{3}(t),\qquad
 \dot{p}^{2}(t)=\beta J^{1}(t)p^{3}(t)-\alpha p^{1}(t)J^{3}(t),\nonumber\\
 \dot{p}^{3}(t)=\beta \big(J^{2}(t)p^{1}(t)-J^{1}(t)p^{2}(t)\big),\nonumber\\
 \dot{J}^{1}(t)=(\alpha-\beta)J^{2}(t)J^{3}(t)+\beta p^{2}(t)p^{3}(t),\qquad
 \dot{J}^{2}(t)=(\beta-\alpha)J^{1}(t)J^{3}(t)-\beta p^{1}(t)p^{3}(t),\nonumber\\
 \dot{J}^{3}(t)=0 \quad  \Rightarrow \quad J^{3}={\rm const}\doteq M, 
 \end{gather*}
where for simplicity we have posed $\alpha\doteq B_{3}^{-1}$ and $\beta\doteq
B_{1}^{-1}$. The new variables suggested by Kirchhof\/f are given by $p^{1}=s\cos(f)$,
$p^{2}=s\sin(f)$, $J^{1}=\sigma\cos(\psi+f)$, $J^{2}=\sigma\sin(\psi+f)$. In
terms of these variables the equations of motion can be written as:
\begin{gather*}
 \dot{s}(\tau)=-\sigma(\tau) p^{3}(\tau)\sin(\psi(\tau)), \qquad
 \dot{\sigma}(\tau)=- s(\tau)p^{3}(\tau)\sin(\psi(\tau)),\nonumber\\
 \dot{p}^{3}(\tau)=s(\tau) \sigma(\tau) \sin(\psi(\tau)),\\ 
 \dot{\psi}(\tau)=M-p^{3}(\tau)\cos(\psi(\tau))\left(\frac{\sigma(\tau)}{s(\tau)}+\frac{s(\tau)}{\sigma(\tau)}\right),\qquad
 \dot{f}(\tau)=\frac{\sigma(\tau)}{s(\tau)}p^{3}(\tau)\cos(\psi(\tau))-\frac{\alpha}{\beta}M, \nonumber
\end{gather*}
where $\tau \doteq \beta t$, $(\,^{\dot{}}\,)=\frac{\partial(\,\,)}{\partial \tau}$.
By using the constraints given by the Casimirs and the integral $H_{1}$,
i.e.\ $2H_{1}=\sigma^{2}+(p^{3})^{2}$, $2C_{2}=s^{2}+(p^{3})^{2}$,
$s\sigma\cos(\psi)+Mp^{3}=C_{1}$, one can readily obtains the equation for the
evolution of~$p^{3}$:
\begin{gather}\label{p3p}
\big(\dot{p}^{3}\big)^{2}=\big(\big(p^{3}\big)^{2}-2C_{2}\big)\big(\big(p^{3}\big)^{2}-2H_{1}\big)-\big(Mp^{3}-C_{1}\big)^{2}.
\end{gather}
At this point Kirchhof\/f notes that this equation is integrable and that one
can obtain by the expression of~$p^{3}$ those for the other
variables, but soon after he passes to consider the special case
$J^{1}=J^{3}=p^{2}=0$. At our knowledge the f\/irst author to integrate this system was
G.E.~Halphen in 1886 \cite{Halphen}, also if some authors gives Kirchhof\/f as
reference for the complete integration of the equation of motions\footnote{See
for an example \cite[page 174]{Lamb}.}. The expression for $p^{3}(t)$ of Halphen is written in
terms of Weierstrass $\wp$ function; it reads as:
\begin{gather}\label{Halphen}
p^{3}(\tau)=\frac{1}{2}\left(\frac{\dot{\wp}\big(\tau+K,\Phi+3\Psi^{2},\Psi^{3}
-\Psi\Phi-\Omega^{3}\big)-\Omega}{\wp\big(\tau+K,\Phi+3\Psi^{2},\Psi^{3}-\Psi\Phi-\Omega^{3}\big)-\Psi}\right),
\end{gather}
where
\begin{gather*}
 \Psi =\frac{2C_{2}+2C_{1}+M^{2}}{6},\qquad
\Omega =\frac{MC_{1}}{2}, \qquad \Phi =4C_{2}H_{1}-C_{1}^{2}.
\end{gather*}
Note that the last two arguments of $\wp$ are not its periods but the
elliptic invariants.
In order to f\/it with the given initial condition
$p^{3}(0)$ it is possible to show that one has to choose the value of $K$ according
to the relation:
\begin{gather*}
\wp\big(K,\Phi+3\Psi^{2},\Psi^{3}-\Psi\Phi-\Omega^{3}\big)=\frac{\big(p^{3}(0)\big)^{2}+\dot{p}^{3}\vert_{t=0}-\Psi}{2}.
\end{gather*}
By the expression (\ref{Halphen}) it isn't simple to see that $p^{3}(\tau)$ is actually bounded. It is however possible
to make plain this point by writing the solution in terms of the roots of~(\ref{p3p}). Let us clarify this statement. The key observation is that if one
looks at the r.h.s.\ of~(\ref{p3p}) as an algebraic
equation for $p^{3}$, so that the equation is a quartic in this variable, then it is
possible to show that it has always four {\emph{real}}
roots. This allows to arrange them in order of crescent magnitude and then to
infer from this fact some properties of the solution of~(\ref{p3p}).
The Casimirs and integrals are f\/ixed if
one f\/ixes the initial conditions, so we can assume that the dynamical
variables in these quantities are specif\/ied by their initial values, say at
$\tau=0$. With this in mind, by posing $p^{3}(\tau)=x$ we rewrite~(\ref{p3p}) as:
\begin{gather}\label{p3palg}
\big(x^{2}-2C_{2}\big)\big(x^{2}-2H_{1}\big)-\big(Mx-C_{1}\big)^{2}=f(x).
\end{gather}
If we can f\/ind f\/ive distinct points where $f(x)$ changes its sign, then the l.h.s.\ of
equation~(\ref{p3palg}) has four real roots. These points are collected in Table~\ref{tabella}.

\begin{table}[t]
\centering\caption{The changes in the sign of $f(x)$.}\label{tabella}
\vspace{1mm}
\begin{tabular}{|c|c|}
\hline
$x$ & $f(x)$\\
\hline \hline
$+\infty$ & $>0$ \\
$\sqrt{C_{2}+H_{1}}$& $\leq 0$\\
$p^{3}(0)$ & $\geq 0$\\
$-\sqrt{C_{2}+H_{1}}$ & $\leq 0$\\
$-\infty$ & $>0$\\
\hline
\end{tabular}
\end{table}

Note that if\/f the initial condition are such that
$J^{1}(0)=J^{2}(0)=p^{1}(0)=p^{2}(0)=0$, then $p^{3}(0)$ is equal to one of the
two points $\pm \sqrt{C_{2}+H_{1}}$. But in this case the four roots are
$x=p^{3}(0)$, $ x = J^{3}(0)-p^{3}(0)$, $x = -J^{3}(0)-p^{3}(0)$, with
$x=p^{3}(0)$ a double root, so that also in this case there are four real
roots. So in general the l.h.s.\ of  (\ref{p3palg}) has four real roots, with \emph{at
  most} three equals and with \emph{at least} one negative (for obvious
reasons we do not consider the trivial case when all the dynamical variables are initially equal to zero). Given the reality
of the roots, it is possible now to sort them in an increasing order of magnitude so that by labelling
with~$a$,~$b$,~$c$,~$d$, we can assume $a \geq b \geq c \geq d$. Note also that
$p^{3}(0)$ lies in the interval $(c,b)$. Equation~(\ref{p3p}) can be written
then as:
\begin{gather}\label{p3pabcd}
(\dot{p}^{3})^{2}=(p^{3}-a)(p^{3}-b)(p^{3}-c)(p^{3}-d).
\end{gather}
The integration of (\ref{p3pabcd}) is reduced to a standard elliptic integral
of the f\/irst kind by the substitution \cite{Armitage}
$z^{2}=\frac{(b-p^{3}(\tau))(a-c)}{(a-p^{3}(\tau))(b-c)}$. After some algebra
we obtain:
\begin{gather}\label{p3}
p^{3}(\tau)=\frac{b-\mu^{2}a \, \snJ (v+\eta\tau,k)^{2}}{1-\mu^{2}\,\snJ (v+\eta\tau,k)^{2}},
\end{gather}
 where
 \begin{gather*}
 \mu^{2} =\frac{b-c}{a-c},\qquad
k^{2} =\frac{(a-d)(b-c)}{(a-c)(b-d)}, \qquad  \eta^{2} =\frac{(a-c)(b-d)}{4},
 \\ v= \snJ^{-1}\left(\sqrt{\frac{(b-p^{3}(0))(a-c)}{(a-p^{3}(0))(b-c)}},k\right).
\end{gather*}
The symbol ``$\snJ$'' denotes the Jacobi elliptic function of modulus~$k$. As can
be seen by the expansion of $p^{3}(\tau)$ in the neighborhood of $\tau=0$, the
sign of $\eta$ has to be chosen according to the sign of $\dot{p}^{3}(0)$,
i.e.\
$\textrm{sgn}(\eta)=-\textrm{sgn}(\dot{p}^{3}(0))=-\textrm{sgn}\big(J^{2}(0)p^{1}(0)-J^{1}(0)p^{2}(0)\big)$.
Note that $p^{3}(\tau)$ is bounded in the set $(c,a)$, that is $c\leq
p^{3}(\tau)\leq a$. Having (\ref{p3}) it is a simple matter to write down
the expressions for the other dynamical variables:
\begin{gather*}
 s(\tau)=\sqrt{2C_{2}-(p^{3}(\tau))^{2}}, \qquad  \sigma(\tau)=\sqrt{2H_{1}-(p^{3}(\tau))^{2}},\qquad
 \cos(\psi(\tau))=\frac{C_{1}-Mp^{3}(\tau)}{s(\tau)\sigma(\tau)},\nonumber\\
 f(\tau)=f(0)+\int_{0}^{\tau} \left(
\frac{s(z)}{\sigma(z)}p^{3}(z)\cos(\psi(z))-\frac{\alpha}{\beta}M \right) dz.
\end{gather*}
The continuous f\/low is in general quasi-periodic. One can ask what are the conditions under which the f\/low becomes periodic. Note that $p^3(\tau)$, $s(\tau)$, $\sigma(\tau)$ and $\cos(\psi(\tau))$ are all periodic functions with the same period (obviously we are interested only in real periods). Such period is given by
\begin{gather}\label{per}
\mathcal{T}= \frac{2K(k)}{\eta},
\end{gather}
and depends only on the initial conditions and not on the parameters $\alpha$ and $\beta$ of the model.
In~(\ref{per}) $K(k)$ is the complete elliptic integral of f\/irst kind~\cite{Armitage}, $k$ and $\eta$ are as given in~(\ref{p3}).
If the function $f(\tau)$ is such that
\begin{gather}\label{periodic}
f(\tau+n\mathcal{T})=f(\tau)+2m\pi, \qquad (n,m)\in \mathbb{Z}^2
\end{gather}
then it isn't dif\/f\/icult to see that the motion is indeed completely periodic. In this case the trajectories of the moving point ($p^1(\tau), p^2(\tau), p^3(\tau)$), constrained on the two-sphere with constant radius given by $r^2=\mathbf{p}(\tau)\cdot\mathbf{p}(\tau)$, close. The condition (\ref{periodic}) is equivalent to the following constraint on the ratio of the parameters $\alpha$ and $\beta$:
\[
\frac{\alpha}{\beta}=\frac{\int_0^{\mathcal{T}} \Big(
\frac{s(z)}{\sigma(z)}p^{3}(z)\cos(\psi(z)) \Big) dz-\frac{2m\pi}{n}}{M\,\mathcal{T}}.
\]

\subsection*{Acknowledgments}
We are grateful to both referees for their constructive comments and criticisms, and in particular to one of them for his crucial remarks and for having brought to our attention the article~\cite{KV}.

The research underlying this paper has been partially supported by the Italian MIUR, Research Project ``Integrable Nonlinear Evolutions, continuous and discrete: from Water Waves downwards to Symplectic Map'', Prot. n. 20082K9KXZ/005, in the framework of the PRIN 2008: ``Geometrical Methods in the Theory of Nonlinear Waves and Applications''.

\pdfbookmark[1]{References}{ref}
\LastPageEnding
\end{document}